\documentclass[12pt]{iopart}
\usepackage{citesort}

\usepackage{amsfonts}
\usepackage{amssymb}
\expandafter\let\csname equation*\endcsname\relax
\expandafter\let\csname endequation*\endcsname\relax
\usepackage{amsmath}
\usepackage{amsfonts}
\usepackage{latexsym}
\usepackage{diffcoeff}
\usepackage{color}
\usepackage[utf8]{inputenc}
\usepackage{tikz-cd}

\newcommand{\marcinFin}[1]{{\color{black} #1}}

\newcommand{\SL}{SL(2,\mathbb C)}
\def\openone{{\rm 1\kern-.22em l}}
\newtheorem{thm}{Theorem}[section]

\newtheorem{lem}[thm]{Lemma}

\newtheorem{df}{Definition}[section]

\begin{document}

\title{On construction of finite averaging sets for $SL(2, \mathbb{C})$ via its Cartan decomposition}

\author{Marcin Markiewicz$^1$ and Janusz Przewocki$^2$}

\address{$^1$International Centre for Theory of Quantum Technologies (ICTQT),
University of Gdansk, 80-308 Gdansk, Poland}
\address{$^2$Institute of Mathematics, University of Gdansk, Wita Stwosza 57, 80-308 Gdansk, Poland}
\ead{marcinm495@gmail.com, jprzew@mat.ug.edu.pl}
\vspace{10pt}


\begin{abstract}
Averaging physical quantities over Lie groups appears in many contexts across the rapidly developing branches of physics like quantum information science or quantum optics. Such an averaging process can be always represented as averaging with respect to a finite number of elements of the group, called a \emph{finite averaging set}. In the previous research such sets, known as $t$-designs, were constructed only for the case of averaging over unitary groups (hence the name \emph{unitary $t$-designs}). In this work we investigate the problem of constructing finite averaging sets for averaging over general non-compact matrix Lie groups, which is much more subtle task due to the fact that the the uniform invariant measure on the group manifold (the Haar measure) is infinite. We provide a general construction of such sets based on the Cartan decomposition of the group, which splits the group into its compact and non-compact components. The averaging over the compact part can be done in a uniform way, whereas the averaging over the non-compact one has to be endowed with a suppresing weight function, and can be approached using generalised Gauss quadratures. This leads us to the general form of finite averaging sets for semisimple matrix Lie groups in the product form of finite averaging sets with respect to the compact and non-compact parts. We provide an explicit calculation of such sets for the group $\SL$, although our construction can be applied to other cases. Possible applications of our results cover finding finite ensembles of random operations in quantum information science and quantum optics, which can be used in constructions of randomised quantum algorithms, including optical interferometric implementations.

\end{abstract}

%

%

%
%
%

\section{Introduction}

\subsection{Averaging over compact and non-compact Lie groups}

The theory of Lie groups lies at the heart of mathematical physics, since it provides a unified framework for describing symmetries in physical systems \cite{Tung}. This includes in particular the space-time symmetries (Galileo group, Lorentz group,  Poincare group) and internal symmetries  (gauge groups for fundamental interactions, groups describing the dynamical symmetries) \cite{Gilmore}. 

One of the crucial issues connected with symmetries is averaging physical quantities over the symmetry operations, which appears in many contexts in several branches of physics: theory of reference frames in quantum physics \cite{Bartlett07}, Gaussian quantum optics \cite{Symplectic00}, \cite{Serafini07} and relativistic quantum communication \cite{Dragan15}, \cite{Piani16} to name a few. Averaging a physical quantity over a Lie group is typically represented as an integral including the averaged quantity transformed by the action of some representation of the group. 
It turns out that such an operation can be always expressed as a finite sum  including only the action of specific group elements, forming a \emph{finite averaging set} \cite{Seymour84}.

The theory of finite averaging sets for symmetry groups has been developed only for the case of unitary groups. Such constructions are known in the literature as \emph{unitary $t$-designs} \cite{Dankert05},\cite{Gross07}, \cite{Scott08}, \cite{Dankert09}, \cite{Roy09}, \cite{Brandao13}, \cite{Webb16}, \cite{Zhu16}, \cite{Hunter19}, \cite{Bae19},  they appear in the context of averaging multiparticle quantum states over a collective action of unitary transformations (meaning that the same transformation acts on each of the particles). 

The case of averaging over unitary groups is peculiar, since these groups are compact, and therefore possess an invariant \emph{finite} Haar measure, which enables to perform the averaging process in a maximally uniform way. Averaging over non-compact groups is a much more subtle task, since the Haar measure is not finite in this case, therefore the maximally uniform averaging would lead to infinite values of the averaged quantities. In order to accomplish this task it is necessary to modify the measure by some suppressing factor, which guarantees finiteness of the results. \marcinFin{As a pedagogical example of this issue one may consider the work \cite{Blume14}, in which a non-existence of finite averaging sets for integrating over Gaussian states in quantum optics is traced back to the divergence of a uniform integral over non-compact symplectic group $Sp(1,\mathbb R)$, representing area-preserving linear transformations on the optical phase space. Note however that the result of \cite{Blume14} \emph{does not imply} the non-existence of finite averaging sets for non-compact symmetry groups if the averaging process is done with respect to properly modified measure, as will be shown further in our work.}

The procedure of averaging over non-compact groups with respect to a modified measure was applied in some physical contexts, like averaging over so called Gaussian operations in quantum optics, represented by the non-compact real symplectic group $Sp(d,\mathbb R)$ \cite{Serafini07}, averaging quantum states over Galileo transformations \cite{Piani16} and Lorentz transformations \cite{Dragan15}, \marcinFin{averaging over infinite-dimensional representations of the non-compact group $SU(1,1)$ in the context of optical quantum state tomography \cite{Carmeli09}}, and averaging over Lorentz-transformed trajectories of the particles in scattering theory \cite{Levy64}. 

\subsection{Main results}
\marcinFin{In this work we state a completely novel problem of constructing finite averaging sets for averaging over \emph{finite-dimensional  tensor representations} of non-compact groups.} We show how to construct such sets for matrix Lie groups basing on the Cartan decomposition of the group. Our construction leads to  finite averaging sets with product structure, namely they are products of finite averaging sets for the maximally compact components of the group and for the maximal abelian subgroup, which is the non-compact part. 

Although this construction is general, its application to concrete examples meets several problems, the most important of which is the direct calculation of the Haar measure of the group in the Cartan decomposition.
In this work we provide an explicit construction of such sets for the group $\SL$, assuming that a proper finite averaging sets for the group $SU(2)$ are known. Our construction allows for applying arbitrary suppressing measure which guarantees finitness of the results, however in general such arbitrary choice demands numerical calculations in order to determine the sets. On the other hand we show one example of choice of such measure, which leads to semi-analytical results. We will refer to finite averaging sets for $\SL$ group as $\SL$-$t$-designs in analogy with the unitary ones.

\subsection{Physical motivation and possible applications}

Our main motivation for this work is the fact that on the one hand the theory of averaging over non-compact groups of transformations is weakly developed across the entire physics, on the other hand such groups represent many important symmetries of the physical systems, like the space-time symmetries (Lorentz group, Poincare group), symmetries of operations in quantum optics (symplectic groups) and symmetries of multiparticle quantum states ($SL(d,\mathbb C)$ groups).

The precise reason for investigating the aspect of finite averaging sets comes from two perspectives. Firstly, the problem is interesting itself from the purely mathematical point of view, since no such construction for the case of averaging over non-compact groups is known. Secondly, finite averaging sets for unitary groups have already found many applications in the context of quantum information science, where the main objects of investigation are correlations between many local subsystems, the most profound of which are known as \emph{quantum entanglement} \cite{HHH09}. Such systems possess important symmetries, namely a class of local operations that \emph{do not increase} the strength of global correlations which arise when locally measuring multipartite quantum systems. Local unitary transformations on the quantum states of many  particles represent such operations which \emph{preserve} the strength of quantum correlations. Finite averaging sets for unitary transformations found numerous applications in the theory of quantum information protocols, mainly as a finite source of local random operations \cite{Gross07}, \cite{Scott08},\cite{Dankert09}, \cite{Roy09}, \cite{Brandao13}, \cite{Gross15}, \cite{Zhu16}, \cite{Bae19}. Now it turns out that the group $SL(d,\mathbb C)$ represents the most general class of local  operations that \emph{do not increase}  global quantum correlations, namely Stochastic Local Operations and Classical Communication (SLOCC) on multipartite states with $d$ degrees of freedom in local subsystems \cite{Dur00}, \cite{Donald02}, \cite{Ver02}, \cite{Avron09}, \cite{Sawicki14}, \cite{Zhang16}. Finite averaging sets with respect to such operations may find applications as finite source of the most general quantum operations not increasing correlations in the theory of general quantum circuits.

Our work provides explicit construction of finite averaging sets for the $\SL$ group only, so it can be directly applied to find finite averaging sets for SLOCC operations on collections of two-level quantum systems (qubits). However, our method of constructing product $t$-designs, augmented with appropriate calculations of the form of Haar measure with respect to the Cartan decomposition, can be directly applied to all other matrix Lie groups, like the groups $SL(d, \mathbb C)$, representing SLOCC operations on collections of $d$-level quantum systems, or real symplectic groups, finite averaging sets of which would represent \emph{finite} ensembles of random transformations for Gaussian quantum optics \cite{Symplectic00}, \cite{Serafini07}.

\subsection{Mathematical issues of the construction}

Our approach to construct finite averaging sets on non-compact groups assumes that the averaging process should be as uniform as possible. In order to fulfill this requirement we decompose the group in a way which separates the compact and non-compact factors of group elements using Cartan decomposition of the group's Lie algebra, which translates to the so-called \emph{KAK} decomposition on the level of Lie group. To the best of our knowledge such approach to averaging has been used only once in the literature for averaging over non-compact real symplectic group $Sp(d,\mathbb R)$ \cite{Serafini07}, however not in the context of finite averaging sets. 

The Cartan decomposition on the level of a Lie group represents the group as a product (cf. Section \ref{sec:Cartan}) of the maximally compact subgroups intertwined with the maximal abelian subgroup which represents the non-compact factor of the group. We show that finite averaging sets over a non-compact Lie group can be constructed as specific products of the averaging sets corresponding to averaging over the compact and non-compact parts separately. The averaging over compact parts is done in a fully uniform way, whereas the averaging over the non-compact part has to be endowed with a suppressing factor which assures finitness of the results. 

In the case of $\SL$ group, the compact factors can be identified with the $SU(2)$ groups, for which the finite averaging sets ($t$-designs) for some values of $t$ are known \cite{Gross07}, \cite{Dankert09}, \cite{Webb16}. In order to construct finite averaging sets over the non-compact part we use the theory of generalised Gauss-type quadratures~\cite{NumericalStoer}. These quadratures give finite averaging sets for generalised polynomial functions of arbitrary degree utilizing a specific functional basis -- the orthogonal polynomials. These polynomials are defined recursively by the demand of orthogonality with respect to a generalised scalar product, which involves the weight function, playing the role of the suppressing factor for the non-finitness of the Haar measure over the entire group. Typically the orthogonal polynomials can be found only numerically, however for some choices of the weight function analytical constructions are known. We show one example of such choice which leads to Laguerre polynomials.

\subsection{Outline and notation}
Our work is organized as follows. In Section \ref{sec:Gen} we formally introduce the notion of finite averaging sets and $t$-designs. Section \ref{sec:Cartan}
is focused on defining the Haar integral over semisimple Lie groups in Cartan decomposition parametrisation together with an explicit calculations for the group $\SL$. Next, Section \ref{sec:Quad} discusses the theory of generalised quadratures based on orthogonal polynomials.
In Section \ref{sec:GenDes} we present the main result of our work, namely general form of finite averaging sets for semisimple Lie groups in a product form based on Cartan decomposition, and provide a constructive procedure of finding $t$-designs for $\SL$ with respect to arbitrary weight function, whereas in Section \ref{sec:LagDes} we give a semi-analytical construction of such a $t$-design with respect to the Laguerre-type weight function. In Section \ref{sec:example} we provide explicit examples of $\SL$-$t$-designs for $t=2,3,5$. 
\marcinFin{Section \ref{sec:SLOCC} is devoted to applications of constructed designs in the context of averaging multipartite quantum states over collective SLOCC operations.}
Final Section \ref{sec:Concs} includes summary of results, discussion of physical applications and some open issues.

Unless specified otherwise, we use the following notational conventions. Sets (including the ones with internal structure, like groups) are denoted by calligraphy letters, e.g. $\mathcal F, \mathcal X$. Matrices are denoted by latin or greek capital letters, like $U, V, \Sigma$. Small greek letters $\mu$ and $\omega$ always denote measures, whereas small fraktur letters, like $\mathfrak g, \mathfrak l$, refer to Lie algebras.

\section{General issues about finite averaging sets and $t$-designs}
\label{sec:Gen}

The concept of a \emph{finite averaging set} has been explicitly defined  by Seymour and Zaslavsky in their seminal paper \cite{Seymour84}. Here, we will use weighted version of their original definition:
\begin{df}
Let $\mathcal{F}=\{f_i\}_{i=1}^m$ be a family of continuous, integrable functions from a path-connected topological space $\mathcal S$ to the space $\mathbb{R}^p$. Let us assume that $\mathcal S$ is equipped with a finite and positive measure $\mu$. A finite averaging set for family $\mathcal{F}$ with respect to real weights $\{w_k\}$ is a finite set of points $\mathcal{X}=\{x_k\} \subset \mathcal S$ which fulfills the following property:
\begin{equation}
    \forall_{i}\,\,\frac{1}{\mu(\mathcal S)}\int_{\mathcal S} f_i(x)d\mu(x)=\sum_{x_k \in \mathcal{X}}w_k f_i(x_k).
\end{equation}
\label{defSZ}
\end{df}
The above definition can be viewed as a generalisation of both the mean value theorem (with respect to a family of functions) and quadrature formulae (being exact for a given class of functions, cf. Section~\ref{sec:Quad}). The main theorem proved in \cite{Seymour84} states that the finite averaging sets with respect to uniform weights $w_i=1/|\mathcal X|$ always exist, however the proof is purely existential. 

The idea of finite averaging sets is related to a geometrical concept of a $t$-design. Such connection appeared for the first time in an implicit form in the work of Delsarte et. all. \cite{Delsarte77} in the context of spherical $t$-designs \cite{Hong82}, \cite{Bajnok91}. Spherical $t$-designs were defined in geometrical terms as finite subsets of points on the $d$-dimensional sphere with the defining property that the sum of monomials of degree at most $t$ evaluated on these subsets is invariant with respect to orthogonal transformations on the set of these points (\cite{Delsarte77}, Definition 5.1). Further in the same work it is shown that this definition is equivalent to a property, that such a sum can be represented as a Haar integral of the monomials over the entire sphere, which in the current terms means that the spherical $t$-designs form a finite averaging sets for polynomial functions defined on a sphere. Subsequently the idea of spherical designs was generalised to unitary designs \cite{Dankert05}, which form a finite averaging sets for polynomial functions defined on manifolds of unitary groups. Furthermore, this concept can be extended to arbitrary matrix Lie groups. Below we provide a definition which is a direct generalisation of the one found in \cite{Roy09} for unitary $t$-designs:

\begin{df}
Let $\mathcal G$ be a matrix Lie group, i.e. there exists a faithful representation of $\mathcal G$ on $\mathrm{Aut}(V)$ -- the group of automorphisms of a finite dimensional vector space $V$. Let $L_g$ denote the linear map on $V$ associated to $g \in \mathcal G$ with respect to the representation and let $\mu$ be a measure on the group manifold which fulfills the conditions from Definition \ref{defSZ}. A finite set $\{(L_i, w_i)\}_i$ of pairs, where $L_i \in \mathrm{Aut}(V)$ are elements  in the image of the representation and $w_i$ are associated weights, is called a finite averaging set of order $t$ ($t$-design on group $\mathcal G$) if:
\begin{equation}
    \frac{1}{\mu(\mathcal G)}\int_{\mathcal G}   L_g^{\otimes t}\otimes(L_g^{*})^{\otimes t}d\mu(g)=\sum_{i}w_i L_i^{\otimes t}\otimes(L_i^{*})^{\otimes t},
    \label{defGT}
\end{equation}
in which the star $^*$ denotes complex conjugate of the matrix.
\label{def:tdesign}
\end{df}

Note that from the representation-theoretic point of view the function that is being averaged in the above definition is a reducible representation of the group $\mathcal G$ on the \marcinFin{\emph{finite-dimensional tensor space}} $V^{\otimes t}\otimes (V^*)^{\otimes t}$. Such representations are used in the tensorial approach to complete characterization of irreducible representations of matrix Lie groups (\cite{Tung}, Chapter 13).

The above matrix definition is equivalent to the original Definition~\ref{defSZ}, if we choose~$\mathcal S$ to be the group manifold of $\mathcal G$, $\mu$ -- a measure on this manifold, and $f_i$ -- real and complex parts of the entries of the matrix $L^{\otimes t}\otimes(L^{*})^{\otimes t}$.

\marcinFin{We point out that the integrals of polynomial functions of elements of \emph{compact} matrix Lie groups (unitary, orthogonal, compact symplectic), similar to the integral present in the defining formula \eqref{defGT}, have been exhaustively studied in \cite{Collins06}, \cite{Collins11}, where closed analytic formulas for such integrals can be found. Nevertheless we emphasize that our aim is not to extend this approach to integration to the domain of noncompact groups, instead we aim at constructing \emph{finite averaging sets} for noncompact groups. }

Let us also provide an equivalent definition, 
which is convenient for applications of $t$-designs in many physical contexts and  more familiar to the Quantum Information Science community. Using the following lemma \cite{BZ06}:
\begin{lem}
Let $\Phi$ be a row-wise vectorization of a matrix. Then the following identity holds for arbitrary square matrices:
\begin{equation}
    \Phi(A.B.C)=(A\otimes C^T).\Phi(B),
\end{equation}
where $^T$ denotes transposition of a matrix and where for the clarity of the formula the dot denotes matrix-matrix and matrix-vector multiplication.
\end{lem}
the defining formula \eqref{defGT} can be expressed in the form:
\begin{equation}
    \frac{1}{\mu(\mathcal G)}\int_{\mathcal G}   L^{\otimes t}.\rho.(L^{\dagger})^{\otimes t}d\mu(L)=\sum_{i\in\mathcal{X}}w_iL_i^{\otimes t}.\rho.(L_i^{\dagger})^{\otimes t},
    \label{defGD}
\end{equation}
where ${}^\dagger$ denotes Hermitian transpose and the formula should hold for arbitrary complex matrix $\rho$ of proper dimensions.
The above definition, although containing redundant matrix $\rho$, has the advantage that the map $\tilde\rho\mapsto  \tilde L^{\otimes t}.\tilde\rho.(\tilde L^{\dagger})^{\otimes t}$, where $\tilde \rho$ is a positive-semidefinite Hermitian matrix of trace $1$ and $\tilde L$ is a normalised $\SL$-matrix, has a physical interpretation of a collective SLOCC transformation of a $t$-partite quantum state $\tilde\rho$. \marcinFin{The possible applications of generalised $t$-designs introduced in Def.~\ref{def:tdesign} for description of a collective SLOCC averaging of quantum states are described in Section~\ref{sec:SLOCC} of this work.}

Note that in the Quantum Information context there is  another notion of $t$-designs, namely \emph{projective $t$-designs} \cite{Ambainis07} and recently introduced \emph{mixed-state $t$-designs} \cite{Czartowski20} which are finite averaging sets for the polynomial functions defined on the manifold of pure and mixed finite dimensional quantum states respectively.

In the previous research the only group $t$-designs in the meaning of Definition \ref{def:tdesign} that have been already discussed are \emph{unitary designs}, for which $G=U(d)$ \cite{Dankert05}, \cite{Gross07}, \cite{Scott08}, \cite{Roy09}, 
\cite{Dankert09}, \cite{Webb16}, \cite{Hunter19} \footnote{Note that most known constructions of unitary designs (see e.g. \cite{Gross07}, \cite{Dankert09}, \cite{Webb16}) are themselves unitary representations of finite groups, like Pauli or Clifford groups \cite{Zhu16}.}. In this case the natural choice of a measure $\mu$ is a Haar measure on the group manifold. This assumption cannot be applied in the context of non-compact Lie groups, since there the Haar measure is not finite. Therefore, in our construction of $t$-designs for $SL(2,\mathbb C)$ we modify the Haar measure in order to make it finite, at the same time trying to keep as much of its symmetry and invariance as possible. To achieve this task we utilize the idea of Cartan decomposition of the Lie algebra, which leads to a convenient factorization of the Haar measure on $SL(2,\mathbb C)$ into factors that depend separately on the compact and non-compact constituents.

\section{Cartan decomposition and the Haar integral}
\label{sec:Cartan}

This section is devoted to the Cartan decomposition and its application to express the Haar integral. We present briefly the general theory and the case of special linear group at the same time. 

\subsection{Cartan decomposition for semisimple Lie groups}

Every Lie group $\mathcal G$ can be approximated by a Lie algebra $\mathfrak{g}$ whose elements are represented by vectors tangent to $\mathcal G$ at the neutral element. In the case $\mathcal G = SL(d, \mathbb{C})$,  the Lie algebra $\mathfrak{g} = \mathfrak{sl}(d, \mathbb{C})$ is usually identified with \marcinFin{$d \times d$} trace-free matrices. 

For every $X \in \mathfrak sl(d, \mathbb C)$, the matrix $\exp(X) \in SL(d, \mathbb C)$, and since the exponent map is surjective on some neighbourood of the identity matrix $X$ is sometimes called an infinitesimal generator of $\exp(X)$. Notice, that there is a different convention that is used in physics according to which $i X$ is called a generator of $\exp(X)$. 

\marcinFin{
Every matrix in $\mathfrak{sl}(d, \mathbb{C})$ can be decomposed into its Hermitian and antihermitian parts. In the case of an arbitrary semisimple \footnote{This, by definition means that the Lie algebra $\mathfrak g$ does not have a solvable ideal} Lie algebra $\mathfrak{g}$ this is precisely the Cartan decomposition
$$ \mathfrak{g} = \mathfrak{k} \oplus \mathfrak{p} $$
where $\mathfrak{k}$ is a Lie subalgebra, but $\mathfrak{p}$ is merely a linear subspace. We shall see later in this section, that there exists a similar decomposition of the Lie group $\mathcal G$ itself. Notice, that in the case $\mathfrak{g} = \mathfrak{sl}(d, \mathbb C)$, we have $\mathfrak{k} = \mathfrak{su}(d)$ and $\mathfrak{p} = i \mathfrak k$. Hence, $\mathfrak{sl}(d, \mathbb C)$ can be thought of as a complexification of $\mathfrak{su}(d)$. 
}



Let us assume that $\mathcal G$ is a semisimple Lie group, which by definition means that the Lie algebra $\mathfrak g$ is semisimple.  There is a subgroup $\mathcal K$ corresponding to $\mathfrak k$. It can be shown \cite[Theorem 6.31]{Knapp} that $\mathcal K$ is compact when the center of $\mathcal G$ is finite and the mapping $\mathcal K \times \mathfrak{p} \to \mathcal G $ given by
\begin{equation}
(k, p) \mapsto k \cdot \exp(p)
\label{eq:diffeo}
\end{equation}
is a diffeomorphism. This can be thought of as a Cartan decomposition on the level of the Lie group. 

Moreover, we can decompose our group even further. Let $\mathfrak{a}$ denote a maximal abelian subalgebra of $\mathfrak{p}$. It can be shown that \cite[Theorem 6.51]{Knapp} 
\begin{equation}
\mathfrak{p}=\bigcup _{k\in \mathcal K}\mathrm {Ad} \,k\cdot {\mathfrak {a}},    
\label{eq:diagonal}
\end{equation}
where $\mathrm {Ad} \,k \cdot $ denotes the adjoint action of $\mathcal K$ on $\mathfrak{g}$, which essentially means the similarity transformation. From that it follows that every element $s \in \mathcal P = \exp(\mathfrak p)$ can be expressed as 
$$s = k^{-1} \cdot  a \cdot k,$$ 
where $a \in \mathcal A = \exp( \mathfrak a)$. Therefore, by \eqref{eq:diffeo} we obtain a decomposition of the form
$$ \mathcal G = \mathcal K \cdot \mathcal A \cdot \mathcal K,$$
which means that every element $g \in \mathcal G$ can be expressed as $k \cdot a \cdot k'$, for some $k, k' \in \mathcal K$ and $a \in \mathcal A$.

Exponents of antihermitian matrices are unitary matrices, hence for $\mathcal G = SL(d, \mathbb{C})$ we have $\mathcal K = SU(d)$. Moreover, we can take $\mathfrak{a}$ to be diagonal traceless real matrices. Notice that the formula \eqref{eq:diagonal} then expresses the fact that Hermitian matrices are diagonalisable.  It is important to underline that the formula is true for semisimple groups $\mathcal G$ and it is well known that $\mathfrak{sl}(d, \mathbb{C})$ fulfills this condition. 

On the other hand, notice that given a matrix $X \in SL(d, \mathbb{C})$ its Cartan decomposition is in fact SVD decomposition 
$$ X = U.\Sigma.V, \qquad U, V \in SU(d), \quad \Sigma \in \mathcal A. $$
Additionaly, we need to stress that the decomposition is non-unique, i.e. there exist matrices $U', V' \in SU(d)$, $\Sigma' \in A$ such that 
$$X = U'.\Sigma'.V'.$$
It is easy to see that the matrix $\Sigma'$ is equal to $\Sigma$ up to a permutation of diagonal elements, and there exists a matrix $W \in SU(d)$ such that $U' = U.W$ and $V' = V.W$. 

The Cartan decomposition can be thought of as a set of non-unique coordinates on $SL(d, \mathbb{C})$. We are aware that SVD decomposition for matrices is conceptually simpler than the abstract Cartan decomposition, however we decided to present the abstract theory here, since it is useful for finding the structure of the Haar measure that we present in the remainder of this section. 

\subsection{Haar measure on semisimple Lie groups}

In the following, we will discuss integrating functions defined on group manifold of $\mathcal G$ with respect to Haar measure expressed in the Cartan coordinates. Our considerations here were inspired by lecture notes \cite{Garrett14}, but we decided to put more details here. \marcinFin{It is important to underline here that even though Cartan coordinates are non-unique (e.g. for $\mathcal{G} = SL(2, \mathbb{C})$, being 6-dimensional, we have the coordinate space $\mathcal K \times \mathcal A \times \mathcal K$ being 7-dimensional) we can can use them to express the Haar integral by treating it as a functional on smooth functions on $\mathcal K \times \mathcal A \times \mathcal K$.}

Every smooth real function $f$ on $\mathcal G$ is naturally identified with a function $\hat{f}$ on the Cartesian product $\mathcal K \times \mathcal A \times \mathcal K$. Then, by ``expressing Haar measure $\mu_{\mathcal G}$ via the Cartan decomposition'' we mean finding a differential form $\omega$ on $\mathcal K \times \mathcal A \times \mathcal K$ such that
\begin{equation}\int_{\mathcal K \times \mathcal A \times \mathcal K} \hat{f} \cdot \omega = \int_{\mathcal G} f \, d\mu_{\mathcal G}.
\label{eq:diff_form}
\end{equation}
for every smooth real function $f$. Because $\mu_{\mathcal G}$ should be both left- and right-invariant, $\omega$ needs to have the form
\begin{equation}
\omega = h \cdot  \omega_{\mathcal K} \wedge \omega_{\mathcal A} \wedge \omega'_{\mathcal K}, \textrm{ where } h \textrm{ is a real function on } \mathcal A.
\label{eq:general_haar_form}
\end{equation}
Here, $\omega_{\mathcal K}$ and $\omega'_{\mathcal K}$ denote differential forms representing Haar measures $\mu_{\mathcal K}$, $\mu_{\mathcal K}'$ on the respective copies of $\mathcal K$, while $\omega_{\mathcal A}$ denotes a form representing a Haar measure on $\mathcal A$.

It is important to notice that since $\mathcal A$ corresponds to an abelian Lie algebra, its group manifold is homeomorphic to the Euclidean space. Moreover, as we shall see in Section \ref{sec:GenDes}, the above representation of the Haar measure is sufficient for Theorem \ref{thm:thm1} to be true, provided a set of elements satisfying \eqref{eq:fin_av_set} can be found, which, as we expect is possible in the most general case e.g. using multidimensional quadratures. That is beyond the scope of this article however, and in the next subsection we focus on the case $d = 2$, which implies~$\mathcal A$ being one-dimensional. 

\subsection{Haar measure on $SL(2, \mathbb C)$}

When $\mathcal G = SL(2, \mathbb C)$, then it is possible to write down analytic form of the function $h$. The subgroup $\mathcal A$ contains matrices $A_r$ of the form
$$ A_r = \begin{pmatrix} e^{r/2} & 0 \\ 0 & e^{-r/2} \end{pmatrix}. $$
We already noticed that the element from $\mathcal A$ occurring in the decomposition of an arbitrary matrix is unique up to a permutation of diagonal elements, therefore we can restrict our considerations to $\mathcal A^+ \subset \mathcal A$ consisting of matrices with parameter $r \geq 0$. Hence, we look for a representation of the Haar measure of the form
\begin{equation}
\omega = h(r) \cdot  \omega_{\mathcal K} \wedge \omega_{\mathcal A} \wedge \omega'_{\mathcal K}, \quad r \geq 0.
\label{eq:haarform}
\end{equation}

Now, we justify that finding the form of $h$ amounts to essentially calculating the Jacobian determinant of some linear mapping. Let us consider a subset $\mathcal U$ containing the neutral element of $\mathcal G$. Translation $\mathcal U'$ of $\mathcal U$ by $A_r$ shall have the same value of the Haar measure. Yet, the preimage of $\mathcal U'$ with respect to the mapping $c: \mathcal K \times \mathcal A \times \mathcal K \to \mathcal G$ defined by
$$c: (K, A, K') \mapsto K . A . K' $$
has different value of the Haar measure on $\mathcal K \times \mathcal A \times \mathcal K$ (i.e. the measure associated with the differential form $\omega_{\mathcal K} \wedge \omega_{\mathcal A} \wedge \omega'_{\mathcal K}$) than the preimage of $\mathcal U$. Hence, the factor $h(r)$ in~\eqref{eq:haarform} is there to compensate this effect. 

In the next step we translate $c^{-1}(\mathcal U')$ to become a neighbourhood of the neutral element in $\mathcal K \times \mathcal A \times \mathcal K$. We do it by taking the image with respect to the mapping being the multiplication of the middle factor by $A_r^{-1}$. The whole process can be depicted in the diagram
$$
\begin{tikzcd}
\mathcal K \times \mathcal A \times \mathcal K \arrow{r}{c} \arrow[swap]{d}{A_r^{-1} \cdot} & \mathcal G \arrow{d}{A_r^{-1} \cdot} \\%
\mathcal K \times \mathcal A \times \mathcal K \arrow{r}{c}& \mathcal G
\end{tikzcd}
$$
Now, we can compare measure of the resulting set with the measure of $c^{-1}(\mathcal U)$. Inverse ratio of these measures is approximately equal to $h(r)$. The smaller diameter of $\mathcal U$, the better is the approximation. In the limit, the ratio depends on the local properties of the mapping $\mathcal K \times \mathcal A \times \mathcal K \to \mathcal G$ defined by
$$(K, A, K') \mapsto A_r^{-1} . K . A_r . A . K'. $$
The above mapping can be approximated \footnote{Since $\mathfrak k \oplus \mathfrak a \oplus \mathfrak k$ is a tangent space to $\mathcal K \times \mathcal A \times \mathcal K$ at the neutral element, we mean here that both maps have the same differential at zero} by its counterpart on the level of Lie algebra, i.e. $\varphi: \mathfrak k \oplus \mathfrak a \oplus \mathfrak k \to \mathcal G$ defined by
$$\varphi: (K, A, K') \mapsto A_r^{-1} . \exp(K) . A_r . \exp(A) . \exp(K'). $$
Recall that in the above formula $K$ and $K'$ are antihermitian traceless matrices and $A$ is a traceless diagonal matrix. The differential of this mapping in zero is a linear map $d\varphi: \mathfrak k \oplus \mathfrak a \oplus \mathfrak k \to \mathfrak g$.

If $\mathcal K \times \mathcal A \times \mathcal K$ and $\mathcal G$ were of the same dimension, then  the measure ratio would be equal to the inverse of the Jacobian  of $d\varphi$. However, here it is 7-dimensional space onto 6-dimensional space being mapped -- this is due to ambiguity of SVD decomposition. Therefore, the value of $h(r)$ should be in fact equal to a minor of the Jacobian matrix. As we shall see below, this value is essentially independent of which column of the matrix is being removed in order to calculate the minor. 

In order to find the form of the Jacobian matrix, let us fix a basis of $\mathfrak k$. We can see that it is generated by Pauli matrices multiplied by the imaginary unit: $i \sigma_x, i \sigma_y, i \sigma_z$. Moreover, the Hermitian part of $\mathfrak g$ is generated by $\sigma_x, \sigma_y, \sigma_z$. Therefore, we fix a basis of 
$$ \mathfrak k \oplus \mathfrak p$$ 
to be $i \sigma_y,  \sigma_x , i \sigma_x, \sigma_y, i \sigma_z, \sigma_z$. And the basis of
$$ \mathfrak k \oplus \mathfrak a \oplus \mathfrak k  $$
is  $i \sigma_x, i \sigma_y, i \sigma_z, \sigma_z  ,i \sigma_x', i \sigma_y', i \sigma_z'$.

\textbf{Remark. } Primed matrices in the above basis denote copies of the respective Pauli matrices that span the other copy of $\mathfrak k$. Let us emphasise that the Jacobian matrix calculated below respects orderings of the above bases. 

Conjugation of $i \sigma_z$ by $A_r$ is again $i \sigma_z$. However, conjugation of $i \sigma_x$ is a linear combination of $i \sigma_x$ and $\sigma_y$. Similarly conjugation of $i \sigma_y$ is a linear combination of $i \sigma_y$ and $\sigma_x$. After calculating coefficients of both linear combinations we obtain the matrix representation of the map $d\varphi$:
$$\begin{pmatrix} 
c & 1 & & & & &  \\
-s & 0 & & & & &  \\ 
 &  & c & 1 & & &  \\ 
 &  & -s & 0 & & &  \\ 
  &  & & & 1 & 1 &  \\ 
 &  & & & & & 1  
\end{pmatrix}  $$
where $c = \cosh(r), s = \sinh(r)$. Let us remark that we decided to put empty places instead of zeros in the above matrix for the sake of its structural clarity. 

Notice that there are two possible values for minors of the highest rank in the above Jacobian matrix. The only nonzero value equals $\sinh^2(r)$ however, and we obtain it after removing 6th or 5th column -- these columns correspond to ambiguity of our parametrisation. Therefore, we see that we can put $h(r) = \sinh^2(r)$, hence the Haar measure is represented by
$$\omega = \sinh^2(r) \cdot \omega_{\mathcal K} \wedge \omega_{\mathcal A} \wedge \omega'_{\mathcal K}.$$

The calculations that we do in the following sections involve generalised Gauss quadratures. These are formulae that are exact for polynomials, and therefore it is convenient for us to have a parametrisation of $A_r$'s with matrix elements being polynomials (or Laurent polynomials precisely, cf. end of Section \ref{sec:Quad}). Hence, substituting $x = e^{r/2}$ we obtain the following parametrisation 
\begin{equation}
A_x = \begin{pmatrix} x & 0 \\ 0 & x^{-1} \end{pmatrix}, \textrm{ for } x \geq 1.
\label{eq:Ax}
\end{equation}
\marcinFin{Knowing that we can take $\omega_{\mathcal A} = dr$, it is easy to calculate the form of the Haar measure in the parametrisation specified by formula \eqref{eq:Ax}:}
\begin{equation}
\omega=2 \left( \frac{x^2 - x^{-2}}{2} \right)^2 x^{-1}  \omega_{\mathcal K} \wedge dx \wedge \omega'_{\mathcal K}.
\label{eq:haar_new}
\end{equation}

\section{Generalised Gauss quadratures and orthogonal polynomials}
\label{sec:Quad}

Generalised Gauss quadratures are a tool needed for our construction of $t$-designs on $SL(2, \mathbb{C})$. They are methods to construct sums of the form
$$ \sum_{k = 1}^n w_k f(x_k), $$
in order to approximate integrals of the following type
$$ \int_a^b f(x) w(x) dx, $$
where $[a, b]$ is a given (possibly infinite) interval, $f$ is an arbitrary function and $w$ is a fixed weight-function. The numbers $x_k$ are called nodes of the quadrature, while $w_k$ are weights of the quadrature. 

\marcinFin{
In its classical form, the Gauss quadrature has $b = -a = 1$ and the weight-function $w(x) = 1$, for all $x$. However, there are also many other cases considered in the literature, including multidimensional quadratures \cite{Li08}\cite[p. 891]{Abramowitz72} and a Lie-theoretic approach \cite{Moody11}, \cite{Hrivnak09}, \cite{Hrivnak16} that is somewhat similar to what we consider in this paper. The main difference is however that in \cite{Moody11}, \cite{Hrivnak09}, \cite{Hrivnak16} integration over domains in the Euclidean space is considered, whereas we focus on the integration over the underlying manifolds of some Lie groups.}


For a particular function it may happen that error of the approximation of the integral by the quadrature formula is exactly equal to zero. Then we say that the quadrature is exact for that function. Later in this section we address the problem of finding a linear space of functions for which the given generalised Gauss quadrature is exact. 

We define the inner product of two functions $f, g$ by
$$ \int_a^b f(x) g(x) w(x) dx $$
A family of orthogonal polynomials with respect to this inner product is defined in the following way. Let $p_0(x) = 1$, for all $x \in (a, b)$, polynomial $p_1$ is defined to be the unique polynomial of degree 1 orthogonal to $p_0$. Similarly, we define $p_2, p_3, ...$ so that degree of $p_k$ is equal $k$. 

The fact follows from $p_k$ being orthogonal to $p_0$
\begin{equation}
\int_a^b p_k(x) w(x) dx = 0, \qquad \textrm{ for all } k > 0.
\label{eq:intorthogonal}
\end{equation}

We define nodes $x_1, x_2, ... , x_n$ of the generalised Gauss quadrature to be the roots of $p_n$. There exist weights $w_1, ..., w_n$ such that (cf. \cite[Theorem 3.6.11]{NumericalStoer})
$$ \sum_{i = 1}^n w_i p_k(x_i) = 0, \qquad \textrm{ for all } 0 < k < n,$$
and
$$ \sum_{i = 1}^n w_i p_0(x_i) = \int_a^b w(x) dx.$$
Notice that from \eqref{eq:intorthogonal} it follows that the above formulae, can be shortly written as
$$ \sum_{i = 1}^n w_i p_k(x_i) = \int_a^b p_k(x)w(x) dx, \qquad \textrm{ for all } 0 \leq k \leq n.$$
In other words, the quadrature formula is exact for orthogonal polynomials $p_0, p_1, ..., p_n$. Furthermore, these polynomials are a basis of the space of all polynomials up to degree~$n$, hence the quadratures are exact for all polynomals in this space.

We can go even further. The degree of exactness for the generalised Gauss quadrature is in fact $2n - 1$, i.e. 
$$ \sum_{i = 1}^n w_i f(x_i) = \int_a^b f(x)w(x) dx, \qquad \textrm{ for all polynomials } f \textrm{ with degree less than } 2n.$$
This can be easily proved using the following two facts:
\begin{itemize}
    \item every polynomial of degree less than $2n$ can be divided by $p_n$ with degrees of quotient and reminder being at most $n-1$,
    \item polynomial $p_n$ is orthogonal to every polynomial of degree at most $n$.
\end{itemize}

\textbf{Remark. } The above construction of orthogonal polynomials and associated quadratures can be generalised to Laurent polynomials \cite{LaurentQuad}, i.e. expressions of the form
$$ \sum_{k = -m}^n a_k x^k. $$
We adopt terminology that $n$ is called \emph{degree} of the Laurent polynomial and $m$ is its \emph{order}.

\section{Construction of t-designs with arbitrary weight-function}
\label{sec:GenDes}

Let $\mathcal G$ be a noncompact semisimple matrix Lie group. Each $G \in \mathcal G$ is identified with a complex matrix (cf. Definition \ref{def:tdesign}), and let $\mathcal K \cdot \mathcal A \cdot \mathcal K$ be Cartan decomposition of $\mathcal G$. In this section we construct sets of matrices that may be called ``product $t$-designs'', since they are products of $t$-designs on $\mathcal K$ with finite averaging sets on $\mathcal A$. 

Because $\mathcal G$ is noncompact, its Haar measure is not finite. To compensate this effect and to make integrals of matrix-elements of our representations convergent we use an appropriate weight-function $w$ whose values depend on the $\mathcal A$-part only. In other words, we assume that measure $\mu$ in Definition \ref{def:tdesign} is associated with the differential form (cf. the general form of the Haar measure \eqref{eq:general_haar_form} discussed in Section \ref{sec:Cartan})
\begin{equation}
w \cdot h \cdot \omega_{\mathcal K} \wedge \omega_{\mathcal A} \wedge \omega'_{\mathcal K}.
\label{eq:mu_general}
\end{equation}

Consider a set $\{(K_\alpha, k_\alpha)\}$  being a $t$-design on $\mathcal K$, and let $\{(A_\beta, a_\beta)\}$ be a finite averaging set on $\mathcal A$, by which we mean that
\begin{equation}
\frac{1}{\mu_{\mathcal A}(\mathcal A)} \int_{\mathcal A}  v(A) A^{\otimes t} \otimes A^{*\otimes t} \cdot \omega_{\mathcal A} = \sum_\beta a_\beta A_\beta^{\otimes t} \otimes A_\beta^{*\otimes t}.
\label{eq:fin_av_set}
\end{equation}
where $v = w \cdot h$ and $\mu_{\mathcal A}$ is the measure associated with $v \cdot \omega_{\mathcal A}$. Since $\mathcal A$ is homeomorphic to the Euclidean space, multidimensional quadratures may be the way to find such finite averaging set. 

In the above setting we prove the following
\begin{thm}
The set $\{(G_i, w_i)\}_{i = \alpha\beta\gamma}$ with $G_i = K_\alpha A_\beta K_\gamma$, $w_i = k_\alpha a_\beta k_\gamma$ is a $t$-design on $G$ with respect to $\mu$ represented by formula \eqref{eq:mu_general}.
\label{thm:thm1}
\end{thm}

\textbf{Proof.} Using the property of the tensor product $$(A_1\otimes A_2).(B_1\otimes B_2).(C_1\otimes C_2)=(A_1.B_1.C_1)\otimes(A_2.B_2.C_2)$$ we get:
\begin{eqnarray}
&& \sum_{i}w_i G_i^{\otimes t}\otimes G_i^{*\otimes t}  =   \\ 
&& \sum_{\alpha,\beta,\gamma}w_{\alpha\beta\gamma} (K_\alpha.A_\beta.K_\gamma)^{\otimes t}\otimes (K_\alpha^*.A_\beta^*.K_\gamma^*)^{\otimes t}  =  \nonumber \\ 
&& \sum_{\alpha,\beta,\gamma} w_{\alpha\beta\gamma} (K_\alpha^{\otimes t}.A_\beta^{\otimes t}.K_\gamma^{\otimes t})\otimes (K_\alpha^{*\otimes t}.A_\beta^{*\otimes t}.K_\gamma^{*\otimes t})  =   \nonumber \\
&&\sum_{\alpha,\beta,\gamma} w_{\alpha\beta\gamma} (K_\alpha^{\otimes t}\otimes K_\alpha^{*\otimes t}).(A_\beta^{\otimes t}\otimes A_\beta^{*\otimes t}).(K_\gamma^{\otimes t}\otimes K_\gamma^{*\otimes t})  =   \nonumber \\
&& \left(\sum_{\alpha} k_\alpha K_\alpha^{\otimes t}\otimes K_\alpha^{*\otimes t}\right).\left(\sum_\beta a_\beta A_\beta^{\otimes t}\otimes A_\beta^{*\otimes t}\right).\left(\sum_\gamma k_\gamma K_\gamma^{\otimes t}\otimes K_\gamma^{*\otimes t}\right). \nonumber
\end{eqnarray}

By our assumptions about $\{(K_\alpha, k_\alpha)\}$ and $\{(A_\beta, a_\beta)\}$, we can express the sums with the corresponding integrals:
\begin{eqnarray}
&&\left(\sum_{\alpha} k_\alpha K_\alpha^{\otimes t}\otimes K_\alpha^{*\otimes t}\right).\left(\sum_\beta a_\beta A_\beta^{\otimes t}\otimes A_\beta^{*\otimes t}\right).\left(\sum_\gamma k_\gamma K_\gamma^{\otimes t}\otimes K_\gamma^{*\otimes t}\right) =  \nonumber\\
&&\frac{1}{\mu_{\mathcal K}(\mathcal K) \mu_{\mathcal A}(\mathcal A) \mu_{\mathcal K}(\mathcal K)}\left(\int_{\mathcal K} K^{\otimes t}\otimes K^{*\otimes t} \cdot \omega_{\mathcal K} \right).\left( \int_{\mathcal A}  v(A) A^{\otimes t} \otimes A^{*\otimes t} \cdot \omega_{\mathcal A} \right). \nonumber \\ & & \left(\int_{\mathcal K} K^{\otimes t}\otimes K^{*\otimes t} \cdot \omega_{\mathcal K} \right) = \nonumber \\
&& \frac{1}{\mu(\mathcal G)} \int_{\mathcal K \times \mathcal A \times \mathcal K} \left( K^{\otimes t}\otimes K^{*\otimes t} \right).\left(  A^{\otimes t} \otimes A^{*\otimes t} \right).\nonumber \\ & & \left( K'^{\otimes t}\otimes K'^{*\otimes t} \right) \cdot v(A) \cdot  \omega_{\mathcal K} \wedge \omega_{\mathcal A} \wedge \omega'_{\mathcal K} = \nonumber \\
&& \frac{1}{\mu(\mathcal G)} \int_{\mathcal K \times \mathcal A \times \mathcal K} (K.A.K')^{\otimes t}\otimes (K^*.A^*.K'^*)^{\otimes t} \cdot v(A) \cdot  \omega_{\mathcal K} \wedge \omega_{\mathcal A} \wedge \omega'_{\mathcal K} =  \nonumber \\
&& \frac{1}{\mu(\mathcal G)} \int_{\mathcal G} G^{\otimes t}\otimes G^{*\otimes t} d\mu. \nonumber
\label{pr2}
\end{eqnarray}
where in the above proof we use a fact that $\mu_{\mathcal K}(\mathcal K) \mu_{\mathcal A}(\mathcal A) \mu_{\mathcal K}(\mathcal K) = \mu(\mathcal G)$. \marcinFin{This fact can be easily seen by taking $f \equiv 1$ in formula:
$$ \int_{\mathcal K \times \mathcal A \times \mathcal K} \hat{f} \cdot v \cdot \omega_{\mathcal K} \wedge \omega_{\mathcal A} \wedge \omega'_{\mathcal K} = \int_{\mathcal G} f \, d\mu $$
which is analogous to formula \eqref{eq:diff_form}.}

\begin{flushright}
$\square$
\end{flushright}

The group $SL(2, \mathbb{C})$ fits in the above setting. Therefore, we can apply Theorem \ref{thm:thm1} to construct $t$-desings for any weight-function $w$. In this case, equation~\eqref{eq:mu_general} reads
$$ 2 w(x) \left( \frac{x^2 - x^{-2}}{2} \right)^2 x^{-1}  \omega_{\mathcal K} \wedge dx \wedge \omega_{\mathcal K}'.$$ 
We construct the $SL(2, \mathbb{C})$-design by taking appropriate design for the grup $SU(2)$ and finding finite averaging set over $\mathcal A$ using the method of generalised Gauss quadratures. Recall that we can restrict ourselves to considering $x \geq 1$ (cf.~equation~\eqref{eq:haarform}). Hence, the finite averaging set over $\mathcal{A}$ in Theorem \ref{thm:thm1} satisfies
\begin{equation}
\frac{1}{\mu_{\mathcal A}(\mathcal A)} \int_1^\infty  v(x) A_x^{\otimes t} \otimes A_x^{*\otimes t} dx = \sum_\beta a_\beta A_\beta^{\otimes t} \otimes A_\beta^{*\otimes t},
\label{eq:fin_av_set}
\end{equation}
where
\begin{equation}
v(x) =  2 w(x) \left( \frac{x^2 - x^{-2}}{2} \right)^2 x^{-1}, 
\label{eq:factorv}
\end{equation}
and $A_x$ is defined in equation \eqref{eq:Ax}. In the next section we give an explicit example of such calculation.

\section{Explicit construction of a t-design on $SL(2, \mathbb{C})$ using Gauss-Laguerre quadratures}
\label{sec:LagDes}

Here we construct concrete example of a $t$-design of the type defined in the previous section. We apply Gauss-Laguerre quadratures in order to find finite averaging sets defined by equation \eqref{eq:fin_av_set}. 

Let us notice that \eqref{eq:fin_av_set} can be rewritten as
\begin{equation}
\int_1^\infty \frac{\psi(x) h(x) w(x)}{\mu_{\mathcal A}(\mathcal A)}  dx =  \sum_\beta a_\beta \psi(x_\beta), 
\label{eq:adesign}
\end{equation}
where $\psi(x)$ is a diagonal entry of $A_x^{\otimes t} \otimes A_x^{*\otimes t}$, that happens to be a Laurent polynomial of degree and order not greater than~$2t$ and $h(x)$ is given by the formula:
\begin{equation}
h(x) =  2 \left( \frac{x^2 - x^{-2}}{2} \right)^2 x^{-1}.
\label{eq:factorv}
\end{equation}
Applying Gauss-Laguerre quadratures requires the integrand of \eqref{eq:adesign} to be transformed to the form 
\begin{equation}
g(x) e^{-x},
\label{eq:gausslaguerreintegrand}
\end{equation}
with $g(x)$ being a polynomial. We see that $h(x)$ is a Laurent polynomial with degree~3 and order 5. Therefore, putting $w(x) = x^{2t+5} e^{-x}$ makes $ \frac{\psi(x) h(x) w(x)}{\mu_{\mathcal A}(\mathcal A)} $ to be of the required form. Consequently, the polynomial $g(x)$ is of degree $4t + 8$. Note that the weight function $w(x)$ depends here on the $t$ parameter of the $t$-design, which is a consequence of choosing specific quadrature model demanding polynomial functions. Using quadratures dedicated to Laurent polynomials  
would remove this dependence.

To choose the appropriate order $n$ of the quadrature, recall that generalised Gauss quadratures are exact for polynomials of degrees up to $2n-1$. Therefore, we must have $2n-1 \geq 4 t + 8$.  From that we obtain $n \geq 2t+5$. 

Now, let us look for a set of points $\{x_\beta\}_{\beta = 1}^n$ and associated weights $\{ v_\beta \}_{\beta = 1}^n$ such that
$$
\int_{1}^\infty g(x) e^{-x} dx = \sum_{\beta = 1}^n v_\beta g(x_\beta). 
$$
Typically Gauss-Laguerre quadratures are defined so that the integral is over $[0, \infty]$ interval. Hence, after making the necessary substitution we see that $x_\beta$'s are such that $x_\beta - 1$ is $\beta$th zero of Laguerre polynomial $L_n(x)$, and the weights are calculated using formula (cf. \cite[25.4.45]{Abramowitz72})
$$
v_\beta = \frac{x_\beta-1}{e (n+1)^2 (L_{n+1}(x_\beta-1))^2}.
$$
Finally, for weights in \eqref{eq:adesign} we have
\begin{equation}
\label{abeta}
a_\beta = \frac{x_\beta^{2t + 5} h(x_\beta) v_\beta}{\mu_{\mathcal A} (\mathcal A)},
\end{equation}
where the measure $\mu_{\mathcal A}$ can be associated with the differential:
\begin{equation}
    d\mu_{\mathcal A}=2\left(\frac{x^2-x^{-2}}{2}\right)^2x^{2t+4}e^{-x}dx.
\end{equation}
The design matrices $A_{\beta}$ in formula \eqref{eq:fin_av_set} are given by:
\begin{equation}
\label{Abeta}
    A_{\beta}=\begin{pmatrix} x_{\beta} & 0 \\ 0 & x_{\beta}^{-1} \end{pmatrix}\,\, \textrm{for}\,\, \beta=1,\ldots,2t+5.
\end{equation}

\section{Example of an $\SL$-$t$-designs for $t=2,3,5$.}
\label{sec:example}
In this section we provide an explicit presentation of the $\SL$-$t$-designs for $t=2,3,5$. Our construction comes in two steps. Firstly, we have to find $SU(2)$-$t$-designs and secondly, we have to construct the designs over the non-compact factors. It turns out that unitary $t$-designs for $t=2,3,5$ can be constructed by taking the unitary representations of the symmetry groups of platonic solids; namely the tetrahedral, octahedral and icosahedral groups respectively \cite{QDice}. Some of these representations contain $U(2)$- instead of $SU(2)$-matrices (e.g. the representation of the octahedral group presented in \cite{BZ06}, p. 564), therefore we need the following lemma:
\begin{lem}
Let $\{(K_i, k_i)\}_{i}$ be a $t$-design on $U(d)$. Then $\left\{\left(\frac{1}{\sqrt[d]{\det(K_i)}}K_i, k_i\right)\right\}_{i}$ is \newline a $t$-design on $SU(d)$.
\label{lemSUD}
\end{lem}

\textbf{Proof}. Let us represent the group $U(1)$ as a subgroup of $U(d)$ by identifying elements of $U(1)$ with matrices of the form $e^{i\theta} \openone$. We observe that
\begin{equation}
U(d) = U(1) \cdot SU(d).
\label{eq:uniary_decomp}
\end{equation}
Indeed, for every matrix $U \in U(d)$ the matrix of the form $e^{-i\theta} U \in SU(d)$ provided that $\det(U)=e^{id\theta}$. From that it follows that every unitary matrix $U$ can be decomposed as a product of a matrix $e^{i\theta} \openone = \sqrt[d]{\det(U)} \openone \in U(1)$ and some matrix $W \in SU(d)$, and we have
\begin{equation}
\label{eq:WeqU}
W^{\otimes t} \otimes W^{*\otimes t}  =\left(\det(U)^{-\frac{t}{d}}\right)\left(\det(U)^{-\frac{t}{d}}\right)^* U^{\otimes t} \otimes U^{*\otimes t}=U^{\otimes t} \otimes U^{*\otimes t},
\end{equation}
since $|\det(U)|=|e^{id\theta}|=1$. Notice, that the decomposition of $U$ is non-unique and therefore the factorisation of $U(d)$ expressed by formula \eqref{eq:uniary_decomp} is not a semi-direct product due to nontrivial intersection of both subgroups. 

However, despite \eqref{eq:uniary_decomp} not being a semi-direct product the Haar measure $dU$ on $U(d)$ can be shown to be a product of the Haar measure $dW$ on $SU(d)$ and the Lebesgue measure $d\lambda$ on the circle (i.e. the Haar measure on $U(1)$). Indeed, let $\mathcal I$ denote the intersection of $U(1)$ and $SU(d)$. It is a finite subgroup consisting of matrices of the form $e^{i\theta} \openone $, where $e^{i\theta}$ is a complex root of unity. Let us consider groups $U(1) / \mathcal I$ and $SU(d) / \mathcal I$. Intersection of both these groups is trivial, therefore its product is in fact a semi-direct product. Moreover, the action of $U(1)$ on $SU(d)$ by conjugation is trivial, therefore $(U(1) / \mathcal I) \ltimes (SU(d) / \mathcal I)$ is in fact a direct product and its Haar measure consequently a product measure. From that, the same property for Haar measure on $U(d)$ follows, since Haar measures on $U(1) / \mathcal I$ and $SU(d) / \mathcal I$ are proportional to $d\lambda$ and $dW$, respectively. 

Next, we conclude
\begin{eqnarray}
&&\frac{1}{\mu_U(U(d))}\int_{U(d)} U^{\otimes t} \otimes U^{*\otimes t} dU =\nonumber\\ &&\frac{1}{\mu_{\lambda}(U(1))\mu_W(SU(d))}\int_{U(1)} d\lambda\int_{SU(d)} W^{\otimes t} \otimes W^{*\otimes t} dW=\nonumber\\ &&  \frac{1}{\mu_W(SU(d))}\int_{SU(d)} W^{\otimes t} \otimes W^{*\otimes t} dW.
\label{eq:right}
\end{eqnarray}

Finally, let us fix $W_i=\left(\det(K_i)^{-\frac{1}{d}}\right)K_i$. Due to \eqref{eq:WeqU} we have:
\begin{equation}
\sum_i k_i W_i^{\otimes t} \otimes W_i^{*\otimes t} =\sum_i k_i K_i^{\otimes t} \otimes K_i^{*\otimes t}.
\label{eq:left}
\end{equation}
The set $\{(K_i, k_i)\}_{i}$ is by assumption a $t$-design on $U(d)$, therefore:
\begin{equation}
\sum_i k_i K_i^{\otimes t} \otimes K_i^{*\otimes t} = \frac{1}{\mu_U(U(d))}\int_{U(d)} U^{\otimes t} \otimes U^{*\otimes t} dU.
\label{eq:tdesign}
\end{equation} 

From \eqref{eq:left}, \eqref{eq:tdesign} and \eqref{eq:right} we see that that the set $\{(W_i, k_i)\}$ is a $t$-design on $SU(d)$. 
 
\begin{flushright}
$\square$
\end{flushright}
As mentioned before $U(2)$-$t$-designs for $t=2,3,5$ can be constructed as finite-dimensional representations of the groups: tetrahedral, octahedral and icosahedral respectively. These representations can be found explicitly by expressing the orthogonal transformations corresponding to symmetry transformations of the regular solids by unitary rotations via standard relations between the $O(3)$ and $U(2)$ group. Such construction can be found in \cite{QDice}. On the other hand one can construct these representations  by  taking the presentation of the group and generating a unitary representation \cite{TriangleGroups}, \cite{Magnus}. We will show this method in the context of the icosahedral group $\mathcal T$. Elements of this group can be generated from the following presentation \cite{Magnus} (Section II.4):
\begin{equation}
    \mathcal T=\langle u,v|u^5=v^3=(uv)^2=1\rangle.
\end{equation}
A unitary representation of this group (in fact this is a projective unitary representation, meaning that the relations in the presentation hold up to a phase factor) can be found by taking \cite{Magnus} (Theorem 2.7):
\begin{eqnarray}
    u\mapsto U&=& \begin{pmatrix} -\varepsilon^3 & 0 \\ 0 & -\varepsilon^2 \end{pmatrix},\nonumber\\
    v\mapsto V&=& -\frac{1}{\sqrt 5}\begin{pmatrix} \varepsilon^3-\varepsilon & 1-\varepsilon^4 \\ \varepsilon-1 & \varepsilon^2-\varepsilon^4 \end{pmatrix},
\label{eq:UIcosi}
\end{eqnarray}
where $\varepsilon=e^{\frac{2\pi i}{5}}$.
By generating all $60$ elements of the group with the above substitution we obtain a $U(2)$-$5$-design, which can be verified by the necessary condition for a set being a unitary design called the frame condition (see e.g. \cite{Roy09}, Theorem 1). 

Let us denote the unitary $t$-designs obtained in this way by $\{\left(K'^{(2)}, \frac{1}{12}\right)\}_i$, $\{\left(K'^{(3)}, \frac{1}{24}\right)\}_i$ and $\{\left(K'^{(5)}, \frac{1}{60}\right)\}_i$, where the weights are constant and equal to the reciprocal of the number of elements of the corresponding set. Now if needed,  using Lemma \ref{lemSUD}, we transform them into the corresponding $SU(2)$-$t$-designs $\{\left(K^{(2)}, \frac{1}{12}\right)\}_i$, $\{\left(K^{(3)}, \frac{1}{24}\right)\}_i$ and $\{\left(K^{(5)}, \frac{1}{60}\right)\}_i$. They are the building blocks for the corresponding $\SL$-$t$-designs.
What is left is the explicit construction of the designs for the non-compact part. We will utilize the methods presented in Section \ref{sec:LagDes} based on Gauss-Laguerre quadratures. Let us denote these sets as $\left(A^{(t)}_{\beta}, a^{(t)}_{\beta}\right)$, in which the matrices $A^{(t)}_{\beta}$ are specified by the formula \eqref{Abeta}, and the corresponding weights by \eqref{abeta}. All these sets contain $2t+5$ elements. Finally the $\SL$-$t$-designs are specified by the sets $\left\{\left(K^{(t)}_iA^{(t)}_{\beta}K^{(t)}_j, \left(k^{(t)}\right)^2a^{(t)}_{\beta}\right)\right\}_{i,\beta,j}$, where the weights $k^{(t)}$ are constants represented by $\left\{\frac{1}{12},\frac{1}{24},\frac{1}{60}\right\}$ for $t=2,3,5$ respectively. Note that the  $\SL$-$t$-designs for $t=2,3,5$ involve respectively $\{1296, 6336, 54000\}$ elements.

\marcinFin{
\section{Finite averaging sets for collective SLOCC operations}
\label{sec:SLOCC}
In this section we discuss one of the possible applications of the $\SL$-$t$-designs to define finite averaging of quantum states over SLOCC operations.
As already mentioned in Section \ref{sec:Gen} the map $\tilde\rho\mapsto  \tilde L^{\otimes t}.\tilde\rho.(\tilde L^{\dagger})^{\otimes t}$, where $\tilde \rho$ denotes arbitrary (mixed) state of $t$ two-level quantum systems and $\tilde L=\frac{L}{||L||}$ is a normalised $\SL$ matrix, can be interpreted as a collective SLOCC transformation of the state $\tilde \rho$. This transformation is non-deterministic, namely succeeds with probability \cite{Avron09}:
\begin{equation}
\label{sucProb}
    p_L(\tilde\rho)=\frac{\operatorname{Tr}(L\tilde\rho  L^{\dagger})}{||L||^2}.
\end{equation}
We propose the following map to encode averaging of a state $\tilde\rho$ over collective SLOCC operations:
\begin{equation}
\label{SLav}
    \tilde\rho\mapsto\frac{1}{\mu(\mathcal G)}\int_{\mathcal G}   \tilde L^{\otimes t}.\tilde\rho.(\tilde L^{\dagger})^{\otimes t}d\mu(L),
\end{equation}
where $\mathcal G$ is fixed to $\SL$.
Although the above map contains subnormalised states, it can be expressed in the following way:
\begin{equation}
\label{SLavNorm}
    \tilde\rho\mapsto\frac{1}{\mu(\mathcal G)}\int_{\mathcal G}  p_{L}(\tilde \rho) \frac{\tilde L^{\otimes t}.\tilde\rho.(\tilde  L^{\dagger})^{\otimes t}}{\operatorname{Tr}\left(\tilde L^{\otimes t}.\tilde\rho.(\tilde  L^{\dagger})^{\otimes t}\right)}d\mu(L),
\end{equation}
and therefore can be seen as an average over properly normalised states acted upon by the SLOCC operations, \emph{weighted} according to the success probability \eqref{sucProb}.

Using our general methods described in previous sections we can construct $\SL$-$t$-designs for the map \eqref{SLav}, which means (in full analogy to \eqref{defGD}):
\begin{equation}
    \frac{1}{\mu(\mathcal G)}\int_{\mathcal G}   \tilde L^{\otimes t}.\tilde\rho.(\tilde L^{\dagger})^{\otimes t}d\mu(L)=\sum_{i\in\mathcal{X}}w_i \tilde L_i^{\otimes t}.\tilde\rho.(\tilde L_i^{\dagger})^{\otimes t}=\sum_{i\in\mathcal{X}}w_i  p_{L_i}(\tilde \rho) \frac{\tilde L_i^{\otimes t}.\tilde\rho.(\tilde L_i^{\dagger})^{\otimes t}}{\operatorname{Tr}\left(\tilde L_i^{\otimes t}.\tilde\rho.(\tilde  L_i^{\dagger})^{\otimes t}\right)},
    \label{defGDSLOCC}
\end{equation}
where in the last step (due to \eqref{SLavNorm}) the process of finite averaging over SLOCC operations can be seen as a finite averaging over properly normalised states weighted according to both: the success probability and the design weight $\omega_i$.

Indeed, since for the Cartan representation of the arbitrary $\SL$ matrix $L$ we have $||L||=||KA_xK'||=x$,
where the matrix $A_x$ is specified in \eqref{eq:Ax}, we can perform all the construction with substitution $L \mapsto \tilde L=K\tilde A_x K'$ with:
\begin{equation}
\tilde A_x = \frac{A_x}{x}=\begin{pmatrix} 1 & 0 \\ 0 & x^{-2} \end{pmatrix}, \textrm{ for } x \geq 1.
\label{eq:AxNorm}
\end{equation}
Utilising the construction from Section \ref{sec:LagDes}, which uses the Gauss-Laguerre quadratures, we need to put the weight function $w(x)=x^{2t+7}e^{-x}$, which leads to the averaging sets of the form:
\begin{equation}
\label{Abeta}
    \tilde A_{\beta}=\begin{pmatrix} 1 & 0 \\ 0 & x_{\beta}^{-2} \end{pmatrix}\,\, \textrm{for}\,\, \beta=1,\ldots,2t+7.
\end{equation}
with respectively modified weights \eqref{abeta}.
}

\section{Conclusions}
\label{sec:Concs}

In this work we investigated the problem of constructing finite averaging sets for averaging over non-compact matrix Lie groups, namely a generalisation of the idea of unitary $t$-designs to the non-unitary case. We have shown that such $t$-designs can be constructed as appropriate products of $t$-designs for the maximal compact subgroups and the maximal abelian subgroups. 
\marcinFin{Such form of the $t$-designs assures entirely uniform averaging over the compact part, shifting all the necessary non-uniformity to the averaging over non-compact component.}
We provided an explicit construction of $t$-designs for the group $\SL$, for which $t$-designs for the maximal abelian subgroup can be constructed using the theory of single-variable generalised Gauss quadratures. Our approach can be applied to other matrix Lie groups, however such application may demand more involved considerations. Firstly, one has to find a form of a Haar measure of the group in Cartan decomposed form, secondly, the maximal abelian subgroup can be multidimensional, which would demand usage of multi-variable generalisations of Gauss-type quadratures, which are not discussed in this work. Nevertheless the proposed method of defining the Haar integral in the Cartan decomposition of the group should in principle work in the general case.

The main possible physical application of our work is a construction of finite ensembles of physical operations represented by matrices from standard Lie groups which would serve as finite averaging sets of operations in implementations of randomised quantum algorithms. This would cover for example finite averaging sets of SLOCC operations on multipartite quantum states, \marcinFin{discussed in Section \ref{sec:SLOCC}}, and Gaussian optical operations in quantum interferometry. 

Our Cartan-decomposition-based approach to finite averaging sets for averaging over Lie groups is adjusted to the case of finite dimensional matrix representations of the groups. The question arises whether this approach can be applied to infinite dimensional unitary representations of such groups \marcinFin{(see e.g. \cite{Chandra47}, \cite{Duc67}, \cite{Conrady11})} which appear in many contexts in physics, especially in relativistic quantum mechanics and quantum field theory. The Theorem of Seymour and Zaslavsky on existence of finite averaging sets \ref{defSZ} does not apply here, since the representation spaces are spanned by countable and not finite families of functions. Nevertheless one may hope that the methods of Cartan decomposition may help in approaching the problem of finding infinite discrete forms of $t$-designs for infinite dimensional representations.

\section*{Acknowledgements}

MM acknowledges his Wife Agnieszka Markiewicz for stating the problem of finite averaging over non-compact groups.

The authors acknowledge discussions and correspondence with Paul Garrett, Marek Ku\'s and Karol \.Zyczkowski.

The work is part of the ICTQT IRAP (MAB) project of FNP, co-financed by structural funds of EU.

\section*{References}
\bibliography{TDesigns}

\end{document}